\documentclass[preprint,aps,tightenlines,floatfix,showpacs]{revtex4}

\usepackage{graphicx}
\usepackage{bm}
\usepackage{amsmath}
\usepackage{amssymb}

\begin{document}
\title{On the  excluded space  in applications of  Feshbach projection
formalism}

\author{S. Karataglidis}
\email{S.Karataglidis@ru.ac.za}
\affiliation{Department of Physics and Electronics, Rhodes University,
P.O. Box 94, Grahamstown, 6140, South Africa.}

\author{K. Amos}
\email{amos@physics.unimelb.edu.au}
\affiliation{School of Physics, University of Melbourne,
Victoria 3010, Australia}

\date{\today}

\begin{abstract}
Various  model applications  in nuclear  structure and  reactions have
been  formulated starting  with the  Feshbach projection  formalism. In
recent  studies a  truncated  excluded space  has  been enumerated  to
facilitate calculation and identify a convergence in expansions within
that  truncation.  However,  the  effect  of  any  remainder  must  be
addressed before results from such can be considered physical.
\end{abstract}


\maketitle

Interest  in the  Feshbach  projection  formalism as  a  way to  unify
reaction  and structure  theory  has increased  in  recent years  with
studies such as those of the no-core shell model (SM), like those made
using $G$-matrix interactions~\cite{Na96}, and  of the continuum shell
model~\cite{Vo06}.  With reaction  theory, the  formalism lies  at the
heart  of   the  recent  multi-channel   algebraic  scattering  (MCAS)
theory~\cite{Am03} though  applications of that  method~\cite{Ca06} so
far  assume that  all important  channels  have been  included in  the
${\cal P}$  space. However, the formalism  has long been  used to make
the     framework    of    the     optical    model     for    elastic
scattering~\cite{Fe58,Fe62}.  In  that use,  all  channels other  than
those of the  elastic scattering are taken to  form the excluded space
$\cal Q$ while the elastic  channels span the space supposedly treated
exactly. The dual space formalism  is elegant but it is almost totally
impractical due to the (usually  large) number of reaction channels to
be  considered in  defining  terms such  as  the dynamic  polarization
potential. Often  then, either a total phenomenological  approach or a
restricted selection of terms of importance in the excluded space need
be  made to facilitate  evaluations. While  the foregoing  centered on
reaction theory, such  is also the case when  the formalism is applied
to consider properties  of an isolated nucleus as  with the $G$-matrix
SM of Navr\'atil and Barrett~\cite{Na96} and with the Continuum  Shell 
Model (CSM) of Volya and Zelelevinsky~\cite{Vo06}. The
question arises  as to  the convergence of  solutions obtained  in the
truncated ($\cal Q$) space to those of the actual complete space.

The dual space form  of the Feshbach theory~\cite{Fe58,Fe62} fragments
the Hilbert space $\{\Psi\}$ into  subspaces ${\cal P}$ and ${\cal Q}$
by the action of projection  operators $P$ and $Q$ respectively on the
space spanned  by the eigenfunctions  $\Psi$ of the  full Hamiltonian.
The subspace ${\cal P}$ is spanned by the functions $P\Psi$ while that
of  ${\cal Q}$  is  spanned by  $Q\Psi$.  The assumption  is that  the
solutions within  $P$ space  can be evaluated  while that of  the full
space ${\cal P}  + {\cal Q}$ cannot. As  projection operators, $P$ and
$Q$ satisfy the conditions
\begin{align}
P^2 & = P;\;\;\; \;\;\; Q^2 = Q;\;\;\; \;\;\; PQ = QP = 0
\nonumber\\
P + Q & = 1 .
\end{align}
Then,  with  the  notation  $H_{XY}   =  XHY$,  where  $X,Y$  are  any
combination  of  the   operators  $P,Q$,  the  complete  Schr\"odinger
equation can be recast as
\begin{align}
\left(E - H_{PP} \right) P\Psi & = H_{PQ} Q\Psi
\nonumber\\
\left(E - H_{QQ} \right) Q\Psi & = H_{QP} P\Psi .
\label{first}
\end{align}
Using the second of these to define $Q\Psi$ in terms of $P\Psi$
then defines the ${\cal P}$ space equation to be solved, namely
\begin{equation}
\left[E - H_{PP} - 
H_{PQ} \frac{1}{\left(E - H_{QQ} \right)} H_{QP}
 \right] P \Psi = 0 .
\end{equation}
Thus the  contribution to  the full Hamiltonian  from coupling  to the
excluded states  (the so-called ``doorway'' states) is  defined as the
additional term
\begin{equation}
\Delta H = H_{PQ} \frac{1}{\left(E - H_{QQ} \right)} H_{QP}.
\end{equation}

If  one  restricts   consideration  first  to  nucleon-nucleus  ($NA$)
scattering, where  the ${\cal  P}$ space may  be taken to  define only
elastic  scattering,  then  $\Delta  H$ is  the  dynamic  polarization
potential aspect  of the optical potential. It  is an energy-dependent
problem  to   evaluate.  At  low   energies,  as  evidenced   by  MCAS
studies~\cite{Am06b},  and  for  light  masses  particularly,  channel
coupling  of the  nucleon  to excited  states  of the  target are  the
essential elements. The DPP that results is highly non-local, complex,
and energy-  and angular momentum  dependent. One consequence  is that
phenomenological  local  potentials  which  are  often  used  are  not
physically  justified. For  such cases  it seems  that the  ${\cal Q}$
space  may be  enumerated with  coupling  of the  incident nucleon  to
relatively  few  states of  the  target  nucleus.  At higher  incident
energies, values for  which the giant resonances of  the target may be
excited,  past studies  have noted  their influence  in  scattering as
doorway states  in second order  processes~\cite{Ge75}. Other channels
which   reflect   in    those   studies   require   residual   complex
(phenomenological) optical  potentials; they should  be nonlocal. When
the available  energy coincides  with the target  in its  continuum, a
successful way to  enumerate ${\cal Q}$ space effects  is to adapt the
KMT  theory~\cite{Ke59}. That  has been  done, for  example,  with the
so-called  $g$-folding   optical  potential  \cite{Am00}.   For  those
energies, one cannot enumerate ${\cal  Q}$ space in any detail but its
effects   in  many   applications  to   date   \cite{Am00,Am06a}  seem
encompassed in  the medium and  Pauli blocking effects defined  in the
infinite  matter  $g$-matrices upon  which  the  $g$-folding model  is
based.

A truncated Hilbert  space is also the key feature  of the shell model
for nuclear structure; being  essential for practical calculation. One
may define the ${\cal P}$ space as that in which the SM interaction is
defined.  The ${\cal  Q}$ space  then  encompasses all  of the  higher
$\hbar\omega$ admixtures lying outside  and which can have effects due
to long range  correlations. It is well known  that by restricting the
${\cal P}$  space to be  just the $0\hbar\omega$ for  valence nucleons
requires  polarization  charges,  typically  of  0.5$e$,  to match the
associated  model results  to  observation. Such  is  a reflection  of
higher  $\hbar\omega$ correlations. However,  by increasing  the basis
size,  e.g. to  encompass  a complete  $(0+2)\hbar\omega$ space,  then
there remain  yet higher  order correlations that  may still  play an
important role.  Whatever the space  enlargement with a SM,  one still
has missing pieces by definition.

The    $G$-matrix   SM   model    interaction   of    Navr\'atil   and
Barrett~\cite{Na96}  as  used  later~\cite{Na98,Na00,Ha03},  considers
coupling up to three-body correlations. In that no-core SM, the ${\cal
  Q}$ space  is assumed to be  made of states including  only those 2-
and 3-body correlations above those  included in the ${\cal P}$ space.
However, then $P + Q \ne 1$ by construction. Higher order correlations
may become  important as the mass  of the nucleus  increases, and with
them, the probability of forming large clusters. $^{12}$C is a case in
point. First  there is  the success of  the $\alpha$-cluster  model in
explaining the  super-deformed $0^+$ state at 7.6  MeV, which suggests
4-particle correlations at  least must be included in  the SM. Second,
it  was observed  that  the  excitation energy  of  the $2^+_1$  state
diverged from the experimental value as the model space was increased.
Finally the  predicted $B(E2;2^+_1 \to  0^+_1)$ is not in  good enough
agreement  with the  measured value.  Even including  a  three nucleon
force,   as  was   done  recently~\cite{Ha03},   did  not   make  much
improvement. By contrast, when  using the (fitted) MK3W interaction in
the $(0+2)\hbar\omega$ model space, very good results for the spectrum
and for the  electron scattering form factors were  found for $^{12}$C
\cite{Ka95}. The action of  the fitting in determining the interaction
implicitly  includes higher-order correlations  not considered  in the
those {\it ab initio}  interactions. While the fitted interactions are
not \textit{ab initio}, the use  of them does illustrate the effect of
the $\cal Q$ space.

The CSM  calculations~\cite{Vo06} allow coupling to  the continuum via
one-  and  two-particle  excitations.  They consider  the  effects  of
coupling to  the continuum (their  restricted ${\cal Q}$  space) above
the standard shell model (${\cal  P}$ space). By so doing they ascribe
widths to resonance  states without changing the energy  of that state
from  what was  found from  a ${\cal  P}$ space  calculation. However,
coupling to  the ${\cal Q}$  space necessarily requires that  there be
contributions  from such  coupling also  to  the real  part of  energy
eigenvalues.  The effects  of higher  order correlations  remains also
unknown for this model.

To consider such limitations made  upon the actual ${\cal Q}$ space of
any  problem,  we make  a  three  space  development of  the  Feshbach
formalism. The  Hilbert space  is divided now  into the ${\cal  P}$, a
reduced,  enumerable,  ${\cal Q}$,  and  a  new  remainder ${\cal  R}$
spaces. There  are three associated projection operators,  $P, Q$, and
$R$ satisfying $P +Q +R =  1$. In this case the Schr\"odinger equation
can be cast into the three projected, coupled equations
\begin{align}
\left(E - H_{PP} \right) P \Psi & =
\left(H_{PQ} Q\Psi + H_{PR} R\Psi \right)
\nonumber\\
\left(E - H_{QQ} \right) Q \Psi & =
\left(H_{QP} P\Psi + H_{QR} R\Psi \right)
\nonumber\\
\left(E - H_{RR} \right) R \Psi & =
\left(H_{RP} P\Psi + H_{RQ} Q\Psi \right) .
\label{second}
\end{align}
These equations reduce to those of Eq.~(\ref{first}) in the limit that
couplings to  ${\cal R}$ space can  be neglected. Such seem  to be the
case for $NA$ scattering at low and intermediate energies. Rearranging
Eqs.~(\ref{second}) gives
\begin{align}
Q\Psi & = \frac{1}{\left(E - H_{QQ}\right)}
\left[H_{QP} P\Psi + H_{QR} R\Psi \right]
\nonumber\\
R\Psi & = \frac{1}{\left(E - H_{RR}\right)}
\left[H_{RP} P\Psi + H_{RQ} Q\Psi \right] .
\label{third}
\end{align}
Using the second  equation for $R\Psi$ in the  first, and rearranging,
gives
\begin{multline}
Q\Psi  = \left[1 -
\frac{1}{\left(E - H_{QQ} \right)} H_{QR}
\frac{1}{\left(E - H_{RR} \right)} H_{RQ}
\right]^{-1} \\
\frac{1}{\left(E - H_{QQ} \right)} 
\left[ H_{QP} + H_{QR} \frac{1}{\left(E - H_{RR} \right)} H_{RP}
\right] P\Psi .
\label{part1}
\end{multline}
Likewise one  can eliminate $  Q\Psi$ from the  Eqs.~(\ref{third}) and
rearrange the result to find
\begin{multline}
R\Psi = \left[1 -
\frac{1}{\left(E - H_{RR} \right)} H_{RQ}
\frac{1}{\left(E - H_{QQ} \right)} H_{QR}
\right]^{-1} \\
\frac{1}{\left(E - H_{RR} \right)} 
\left[ H_{RP} + H_{RQ} \frac{1}{\left(E - H_{QQ} \right)} H_{QP}
\right] P\Psi .
\label{part2}
\end{multline}
Then substituting Eqs.~(\ref{part1})  and (\ref{part2}) into the first
of Eqs.~(\ref{second}) gives the ${\cal P}$ space equations
\begin{multline}
\left\{ E - H_{PP} -
H_{PQ} \left[1 - \frac{1}{\left(E - H_{QQ} \right)} H_{QR}
\frac{1}{\left(E - H_{RR} \right)} H_{RQ} \right]^{-1}
\right.\\
\frac{1}{\left(E - H_{QQ} \right)} 
\left[ H_{QP} + H_{QR} \frac{1}{\left(E - H_{RR} \right)} H_{RP}
\right] \\
- H_{PR} \left[1 -
\frac{1}{\left(E - H_{RR} \right)} H_{RQ}
\frac{1}{\left(E - H_{QQ} \right)} H_{QR}
\right]^{-1}\\
\left.
\frac{1}{\left(E - H_{RR} \right)} 
\left[ H_{RP} + H_{RQ} \frac{1}{\left(E - H_{QQ} \right)} H_{QP}
\right] 
\right\} P\Psi  = 0.
\label{final}
\end{multline}

Again  for  situations where  coupling  to  ${\cal  R}$ space  can  be
ignored, ($H_{PR}  = H_{QR} = 0$),  this reduces to the  form given in
Eq.~(\ref{first}).  If  expansion  is  restricted to  third  order  in
couplings, each inverse bracket  term in Eq.~(\ref{final}) becomes the
unit operator so that a reduced ${\cal P}$ space equation is
\begin{multline}
\left\{ E - H_{PP} - H_{PQ} \frac{1}{\left(E - H_{QQ} \right)} H_{QP} 
- H_{PR} \frac{1}{\left(E - H_{RR} \right)} H_{RP} 
\right. \\
-
H_{PQ} \frac{1}{\left(E - H_{QQ} \right)} 
H_{QR} \frac{1}{\left(E - H_{RR} \right)} H_{RP}
\\
- \left.
H_{PR} \frac{1}{\left(E - H_{RR} \right)} 
H_{RQ} \frac{1}{\left(E - H_{QQ} \right)} H_{QP}
\right\} P\Psi = 0 .
\end{multline}

Thus the coupling  to excluded space of a problem  may be via coupling
to either  the ${\cal Q}$ or  ${\cal R}$ subspaces at  second order in
the expansion as  well as via the chains ${\cal  P} \to {\cal Q}({\cal
  R}) \to {\cal R}({\cal Q}) \to  {\cal P}$ in third order. The states
in ${\cal R}$ space more than  play the role of hallway states in past
studies, for we now allow  for states therein that directly connect by
couplings omitted  in the enumerated  ${\cal Q}$ space. To  the extent
that coupling to  those states in ${\cal R}$  space is not negligible,
such have effect even at  second order in expansions. Thus, unless one
can ensure all important  couplings are collected within an enumerated
${\cal Q}$  space, then  no matter  how much, and  to what  order, the
effects of the enumerated ${\cal  Q}$ space are taken, errors arise at
the second  order level. Results  of using the  $G$-matrix interaction
developed by Navr\'atil and Barrett~\cite{Na96} in large space no-core
shell model  calculations for systems (such  as $^6$Li~\cite{St05} and
$^{12}$C~\cite{Na00,Ha03})      which     exhibit      features     of
$\alpha$-clustering, and  so of four-body  correlations, give evidence
of  such  omission.  That   approach  does  show  convergence  to  the
enumerated ${\cal  Q}$ space complete values, but  those values remain
far  from the  empirical ones.  The  problem starts  at second  order.
Indeed, neglect  of the higher  order correlations in  that $G$-matrix
interaction   has    already   been   identified    as   a   potential
problem~\cite{St05}. In  contrast, the earlier  interactions developed
by Zheng~\textit{et  al.}~\cite{Zh95} seem to  accommodate such higher
order  effects through  the way  they  form the  $G$-matrices, as  the
values of the $B(E2)$ in  $^{12}$C change from 5.45 to 6.9 $e^2$fm$^4$
for  evaluations   in  a  $0\hbar\omega$  and   a  (complete)  no-core
$(0+2)\hbar\omega$  shell model,  respectively. The  results, obtained
using the  OXBASH shell  model code~\cite{Ox86}, are  reasonable given
that the measured value is 7.77 $e^2$fm$^4$. But we note that in other
studies~\cite{Mi92}, made  using a  standard shell model,  problems of
only including  $2\hbar\omega$ excitations were  found. In particular,
the  neglect of excitations  above the  $2\hbar\omega$ level  forces a
much  deeper binding  energy for  the  ground state.  Addition of  the
$4\hbar\omega$ excitations, in part, resolved the problem, as has been
illustrated in the case of $^{16}$O \cite{Ha90}.

While a formalism  with which the effects of  an enumerable ${\cal Q}$
space  can be  accounted may  be  aesthetically pleasing,  it must  be
remembered at what level of evaluation problems can occur when seeking
to apply same to reality. Certainly  the approach is built upon a good
theoretical base  and is an  excellent mathematical exercise.  But for
use  in   considering  real  systems,   it  is  not   necessarily  nor
sufficiently complete  even at second order. That  may be particularly
problematic for deformed nuclei.  The object of structure and reaction
theory is to determine as  much information as possible regarding real
nuclei and in  approaches built upon a Feshbach  formalism, all strong
coupling features need to be  accounted at whatever order they appear.
In principle there  may no solution to this  problem built from theory
made using first principles and one may only have the specification of
effective forces  (phenomenological or otherwise)  to enable structure
and reactions  to be dealt with in  any form of unified  way. It would
seem that  the most fruitful  way to proceed  is to seek  the dominant
states in the spaces to which to couple.

\bibliography{Rev-PQRpaper}

\end{document}